\def \Mpc {h^{-1}{\rm Mpc}}
\def \kpc {h^{-1}{\rm kpc}}

\def \kms {{\rm km~s}^{-1}}
\def \msun {h^{-1} M_\odot}
\def \beqn {\begin{equation}}
\def \eeqn {\end{equation}}
\documentstyle[12pt,aaspp4]{article}

\begin{document}

\title{X-ray Emitting Groups in the Infall Region of Abell 2199}

\author{K. Rines\altaffilmark{1}, A. Mahdavi\altaffilmark{1},
M.J. Geller\altaffilmark{1}, A. Diaferio\altaffilmark{2},
J.J. Mohr\altaffilmark{3,4}, and G. Wegner\altaffilmark{5}}
\authoremail{krines@cfa.harvard.edu}

\altaffiltext{1}{Harvard-Smithsonian Center for Astrophysics, 60 Garden St,
Cambridge, MA 02138 ; krines, amahdavi, mgeller@cfa.harvard.edu}
\altaffiltext{2}{Universit\`a degli Studi di Torino,
Dipartimento di Fisica Generale ``Amedeo Avogadro'', Torino, Italy; diaferio@ph.unito.it}
\altaffiltext{3}{Departments of Astronomy and Physics, University of 
Illinois, 1002 W. Green St. Urbana, Il  61801; jmohr@astro.uiuc.edu}
\altaffiltext{4}{Chandra Fellow}
\altaffiltext{5}{Department of Physics and Astronomy, Dartmouth College, Hanover, NH 03755; gaw@bellz.dartmouth.edu}

\begin{abstract}

Using a large redshift survey covering 95 square degrees, we
demonstrate that the infall region  
of Abell 2199 contains Abell 2197, one or two X-ray emitting groups,
and up to five additional groups identified in redshift surveys.  Our
survey   
shows that the X-ray emitting systems, located at projected radii of
$1.^\circ4, 
1.^\circ9$, and $5.^\circ1$ (2.2, 3.1, and $8.0~\Mpc$), are connected
kinematically to A2199.  A2197 is itself an optically rich cluster;
its weak X-ray emission suggests that it is much less massive than
A2199. The absence of a sharp peak in the infall pattern at the
position of A2197 supports this hypothesis. 
The outermost group is well outside the virial region of A2199 and
it distorts the infall pattern in redshift space.
The two X-ray emitting groups are roughly colinear,
suggesting the existence of an extended ($8.0~\Mpc$) filament.   The
identification of these infalling groups provides direct support of
hierarchical structure formation;  studies of these systems will
provide insights into structure evolution.  Groups in the infall
regions of nearby clusters may offer a unique probe of the physics of
the warm/hot ionized medium (WHIM) which is difficult
to observe directly with current instruments.

\end{abstract}


\section{Introduction}

Hierarchical structure formation predicts that rich clusters of
galaxies should be surrounded by less massive, infalling clusters or
groups. Many authors have discussed substructure within the virial
region (e.g., \cite{gb}; \cite{ds}; \cite{xarch}; \cite{mfg}). 
Reisenegger et al.~(2000, hereafter RQCM) recently demonstrated
dynamical connections 
between many clusters in the Shapley Supercluster and A3558, the
central cluster.  We describe similar observations of the infall region
around A2199 (\cite{abell}) and show that it contains A2197,
one or two X-ray emitting groups, and up to five additional groups
identified in redshift surveys, all of which lie outside the virial
region of A2199.

 In redshift space, 
the infall regions of clusters form a characteristic trumpet-shaped
pattern (bounded by caustics) as galaxies fall into the cluster as the
cluster potential overwhelms the Hubble flow (\cite{kais87};
\cite{rg89}; \cite{dg}).  Galaxies
outside the caustics are outside the infall region; within the
caustics, there are few interlopers.
Here, we use the observed caustics around A2199 to identify members of
the infall region and to compare the structure of the supercluster as
seen in galaxies to the X-ray structure revealed by the {\em ROSAT}
All-Sky Survey (RASS\footnote{The RASS is available at
http://www.xray.mpe.mpg.de/cgi-bin/rosat-survey/ };
\cite{rass}).  We use the X-ray data to show that the infall region
contains bound systems destined for accretion.
The physical scale at the redshift of the supercluster
($cz=9156~\kms$) is $1^\circ = 1.53~\Mpc$ ($H_0 = 100 h~\kms, \Omega _m
= 0.3, \Omega _\Lambda = 0.7$).

\section{Observations}

We have collected 1216 redshifts (957 new or remeasured) in a large
region (95 square degrees) surrounding 
A2199/A2197.  Two of us (JJM and GW) used the Decaspec
(\cite{decaspec}) at the 2.4-m MDM telescope on Kitt Peak to obtain
249 (248 unique)
redshifts of galaxies in the central $1^\circ$ of this region for
a separate Jeans' analysis of the central region of A2199 (J.J.~Mohr
et al.~in preparation).  

We used the FAST spectrograph (\cite{f98}) on the 1.5-m
Tillinghast telescope of the Fred Lawrence Whipple Observatory (FLWO)
to obtain 684 spectra of galaxies within $5^\circ$.5
($\approx$8.4$~\Mpc$) of the center of A2199. The observations are
similar to those described in Rines et al.~(2000) and will appear in
K.~Rines et al.~(in preparation). 
We observed infall galaxy candidates in two campaigns. We selected
targets from digitized images of the POSS I from the Automated Plate
Scanner$\footnote{The APS databases are supported by the
        National Aeronautics and Space
        Administration and the University of
        Minnesota, and are available at
        http://aps.umn.edu/}$. We initially
selected galaxies using the automatic classification system of APS; we
visually inspected these targets to eliminate stars. The first
campaign yielded a deep sample in the central $4^\circ \times 4^\circ$
region around A2199. This sample is complete to 103aE magnitude
$E<16.5$ (roughly equivalent to R$<$16.0) and consists of 305 redshifts. 
The second campaign (379 redshifts) was a shallower survey ($E<16.1$ 
or $R\lesssim 15.6$) of all galaxies within 
$5.^\circ 5 \approx 8.4~\Mpc$ of A2199. The completeness limits are
not very precise because the magnitudes come from multiple plate scans and
because we could not obtain redshifts for some low surface brightness
galaxies.  We include 84
redshifts associated with the groups NRGs385 and NRGs388 obtained by
FAST for a separate study of the X-ray and optical properties of
groups of galaxies (59 published in \cite{andi}; \cite{rasscals},
hereafter MBGR).

We collected the remaining 200 redshifts from ZCAT$\footnote{Available
at http://cfa-www.harvard.edu/$\sim$huchra/zcat}$ and/or the NASA/IPAC
Extragalactic Database$\footnote{The NASA/IPAC Extragalactic Database
is available at http://nedwww.ipac.caltech.edu/index.html}$.
X-ray data for these systems are available from the RASS.  

\section{Defining the Infall Region with Caustics}

\begin{figure}
\figurenum{1}
\plotone{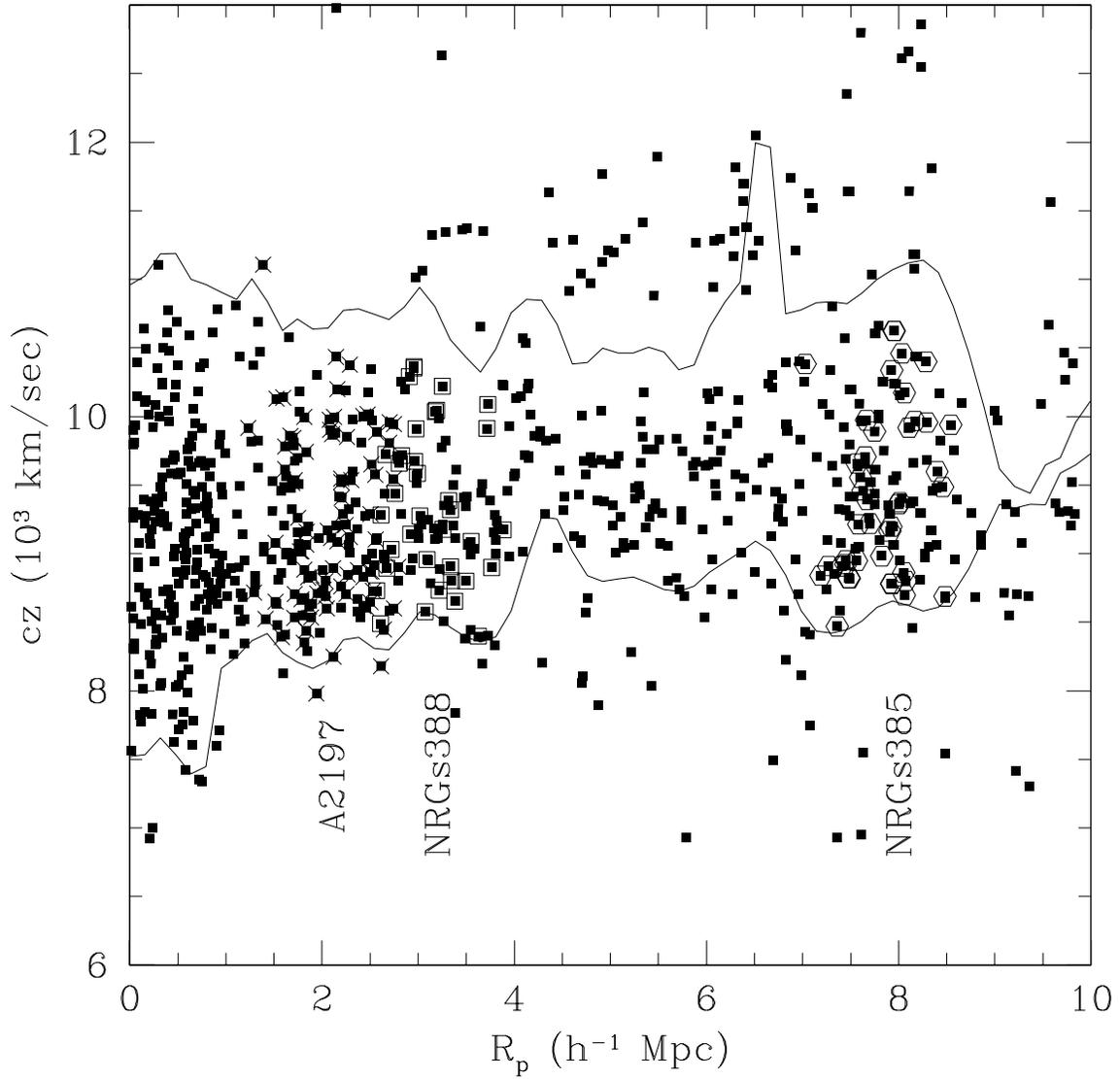} 
\caption{\label{caustics} Projected radius versus redshift for
galaxies surrounding 
A2199.  Lines indicate our estimate of the caustics.  Crosses,
hexagons, and open squares indicate galaxies in A2197, NRGs385, and NRGs388
respectively. There are 743 galaxies within the caustics.}
\end{figure}

Figure \ref{caustics} displays the projected radii and redshifts of
galaxies surrounding A2199. The expected caustic pattern is easily
visible; we calculate the shape with the technique described in
Diaferio (1999) using a smoothing parameter of $q=25$.  We use a
hierarchical clustering analysis to determine the center of the
largest system, A2199; we use this as the center
of the supercluster.  The caustics shown in Figure
\ref{caustics} do not include the symmetry and first derivative
constraints of Diaferio (1999).  We will determine the caustics and
the supercluster mass profile in more detail in Rines et al.~(in
preparation).  Preliminary results indicate that the total
supercluster mass is in the range $5-10 \times 10^{14}~\msun$ within a
radius of $\sim 8~\Mpc$.  For comparison, Markevitch et al.~(1999,
hereafter MVFS) use X-ray data to estimate a mass of $1.7 \times
10^{14}~\msun$ within $0.6~\Mpc$ for A2199.  The A2199 supercluster is
located within the Great Wall (\cite{gh}; Figure 6b of \cite{uzc}),
which may complicate the interpretation of the dynamics of the system.   

\begin{figure}
\figurenum{2}
\label{skyplot}
\plotone{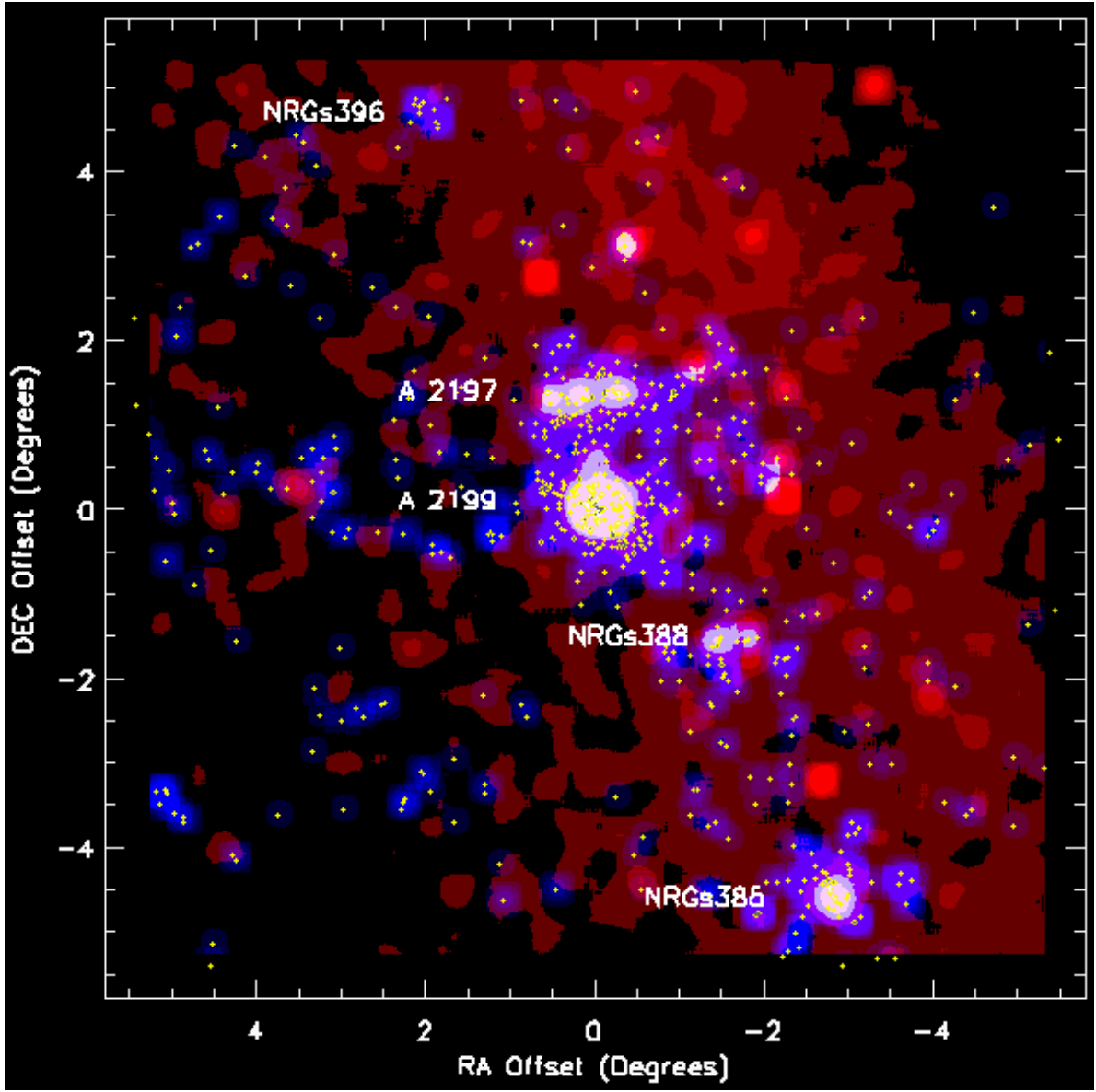} 
\caption{Sky positions of 743 galaxies in the infall region. Red shows X-ray
intensity from the RASS; blue shows smoothed infall galaxy
density.  Regions with X-ray emission and large galaxy density thus
appear purple. Several (most likely 
background) X-ray point sources are present and appear as red
circles. Some groups of galaxies have no associated X-ray
emission detectable in the RASS.} 
\end{figure}

The sky positions of galaxies in the infall region (crosses in Figure
\ref{skyplot}) show several groups in addition to the main cluster. 
Table \ref{properties} lists the coordinates and basic properties of these 
systems (MBGR).   Figure \ref{skyplot} 
displays X-ray contours from the RASS (red) and contours of
the local galaxy density overlaid (blue).  Purple regions in the figure
therefore show regions with both significant galaxy overdensities and
X-ray emission; X-ray emission confirms that a system is bound.
The largest purple region is A2199, at least one
background X-ray source and A2197 are to the N, and NRGs388 and
NRGs385 are to the SW of A2199.  The X-ray emission from A2199  
is quite symmetric relative to other clusters (\cite{mefg};
\cite{maxim}), which suggests that the inner region of A2199 has not
undergone any recent major mergers. The infalling groups are all located
at projected radii significantly larger than the virial radius
($R_{vir} \approx 1.6~\Mpc$, \cite{g98}; $R_{vir} \approx 1.8~\Mpc$,
\cite{maxim}). We identify galaxies within $1~\Mpc$
of these systems in Figure \ref{caustics}.

NRGs385, NRGs388, and A2199 are roughly colinear (Figure
\ref{skyplot}). This alignment may be coincidental or it may indicate
the presence of a filament of galaxies and/or dark matter. A2197 and
an optical group with no X-ray counterpart in the RASS lie roughly
along the extension of this line to the NE.  The apparent X-ray excess
in a NE-SW band, coincidentally along the broad apparent filament, is probably
an artifact of the RASS scan pattern.  Archival pointed PSPC
observations indicate that the X-ray surface brightness of the
filament is less than $4\times 10^{-16}$erg cm$^{-2}$s$^{-1}$arcmin$^{-2}$.
Briel \& Henry (1995) obtain similar limits for other clusters. 

\section{Infalling Systems}

\subsection{Abell 2197}

Rood (1976) first suggested that, due to their close proximity on the
sky and in redshift space, A2197 and A2199 are part of a supercluster.
Gregory \& Thompson (1984) analyzed the pair as a binary cluster.
A2197 is an optically rich Abell cluster with an apparent velocity dispersion
comparable to A2199 (\cite{g98}), but X-ray data suggest that A2197 is
significantly less massive than A2199 (e.g.,
\cite{wjf97}; \cite{jf99}). The absence of a sharp spike in the
caustic pattern at the projected radius of A2197 (Figure 1) indicates
that the infall pattern is dominated by A2199 and that the apparently large
velocity dispersion of A2197 is due in part to its proximity to A2199.

A2199 and A2197 provide an interesting contrast. A2199 is a rich,
regular cluster with both a centrally concentrated galaxy distribution
and X-ray emission. The galaxies around A2197 are significantly
elongated to the SE and NW and they are much
less centrally concentrated than in A2199.  Archival ROSAT data
reveals that the X-ray emission has at least three components (all
three are evident in Figure \ref{skyplot}). The central component is
centered on NGC 6173, the more luminous component to the W is centered
on NGC 6160, and the source to the E is not centered on any bright
galaxies and is probably a background source (C. Jones 2000, private
communication; note 
also that a point source is evident between A2199 and A2197).
The galaxy distribution in A2197 is quite different from the X-ray
emission, though not as dramatically as A754 (\cite{zz}).  We suggest
that A2197 is either forming from the merger of two groups or that it
is being disrupted by A2199. 

\subsection{NRGs388}

The X-ray emission from NRGs388 is centered on a bright elliptical
which dominates the group (see Figure 6 of MBGR).  There is
no significant 
substructure in either the X-ray emission or the galaxy distribution
within NRGs388.  The apparent absence of ram pressure stripping or
tidal distortions may constrain the distribution of intercluster gas
or the shape of the gravitational potential of the supercluster.

\subsection{NRGs385}

NRGs385 is located $5.^\circ1 = 8.0~\Mpc$ or $\sim 5$ virial
radii from A2199.  Although other investigators have found X-ray
emitting groups around rich clusters (e.g., \cite{bhb}; \cite{wcb};
\cite{kb}; \cite{hank}; \cite{rqcm}), to our knowledge no other X-ray
group so far outside the virial radius of a cluster has been linked
to it kinematically.  While the kinematic connection is clear, NRGs385
may not be bound to the supercluster (Rines et al.~in preparation). 

NRGs385 is sufficiently massive that it distorts the caustic pattern
in its vicinity (Figure \ref{caustics}). This effect is expected;
subclustering increases the amplitude of the caustics at all radii
(\cite{d99}), but spikes in the amplitude of the caustics can reveal the
presence of massive subclusters. The existence
of this distortion confirms the hypothesis that subclusters can alter
the sharp caustics expected from spherical infall. It is striking that
the expected distortion is evident from this group but no such distortions
are evident in the infall region of A3558 (see Figure 1 of
\cite{rqcm}) even though it contains infalling clusters which are
probably more massive than NRGs385. The reason for this difference is
unclear, although it might depend on the galaxy population included by
the selection criteria of RQCM.

The peak of the X-ray emission in NRGs385 is offset (at roughly the
3$\sigma$ level) from the
center of the galaxy distribution. The X-ray center is located
$\approx 218~\kpc$ SW of the optical center, away from A2199.  If this
offset is not due to a projection effect, 
it could indicate that either the galaxies do not trace the gravitational
potential of the group or the group has recently undergone a merger.
We rule out the possibility of ram pressure stripping (e.g.,
\cite{gg}) by the warm/hot ionized medium (WHIM) because the required gas
density in the WHIM would produce X-ray emission above our upper limit
on the surface brightness of any filament ($\S 3$).

The shape of the X-ray contours may provide information about the
physical processes occuring in the infalling groups. NRGs385 is
visibly elongated in the direction of A2199 and along the possible
filament; the width of the $10-\sigma$ contour is approximately $26' =
720~\kpc$ NE-SW and approximately $19' = 525~\kpc$ SE-NW (see Figure 6
of \cite{andi}). This elongation may be caused by tidal effects and/or
the presence of a filament.

\subsection{X-ray Faint Groups} 

Several small knots of galaxies are evident in Figure \ref{skyplot}.
Many of these are probably chance superpositions rather than physical
groups.  Four of them are contained in the RASSCALS optical catalog
(MBGR), though none has significant extended X-ray
emission (we list upper limits from the RASS in Table
\ref{properties}).  A fifth possible group, NRGs396 (\cite{rpg};
named according to the convention of MBGR), is located $5^\circ.0 =
7.6~\Mpc$ NE of A2199, but is too poor (4 group members in the CfA
redshift survey) for inclusion in the RASSCALS catalog (MBGR require a
minimum of 5 group members in the CfA redshift survey).  The X-ray
faint group NRGs389 lies between NRGs385 and NRGs388 and is aligned
with the possible filament. 

We find marginal evidence ($2.7\sigma$) of X-ray emission near
NRGs396 in the RASS; 
pointed observations of these groups with {\em XMM-Newton} or {\em 
Chandra} would place tighter constraints on their X-ray properties.
Like NRGs385, NRGs396 and NRGs400 are sufficiently distant from
the center of the supercluster that they may not be gravitationally
bound. 

\section{Discussion}

The infall pattern in redshift space (delineated by the caustics)
indicates that A2197 and one or two X-ray groups (NRGs388 and NRGs385) are
connected kinematically to A2199.  The three systems are located at
projected radii of $1.^\circ4, 1.^\circ9$, and $ 5.^\circ1$ (2.2, 3.1,
and $8.0~\Mpc$) respectively; we also identify five X-ray faint groups
in the infall region.  All three systems with X-ray 
emission are significantly less massive than A2199. 

To our knowledge, NRGs388 (and possibly NRGs385) is the first X-ray
group demonstrably bound to a cluster even though it
lies well outside the virial radius (\cite{rqcm} demonstrate similar
connections between X-ray clusters in the Shapley supercluster).  The 
identification of caustics delineating the infall regions of clusters
reveals the kinematic connections between clusters and distant,
infalling subclusters. 

At large radii from rich clusters, it is difficult to observe the WHIM
directly with current X-ray instruments (\cite{kb}; \cite{pbg};
\cite{co}).  The presence of hot gas in at least some infalling groups 
shows that the intracluster medium in clusters of galaxies can
be accreted from infalling groups.  Like groups at smaller radii
(White et al.~1993, \cite{hank}), these groups may interact with the WHIM
and thus yield insight into its properties and processes such as
cooling and feedback (Pierre et al.~2000).

\acknowledgements

This project would not have been possible without the assistance of
Perry Berlind and Michael Calkins, the remote observers at FLWO, and
Susan Tokarz, who processed the spectroscopic data.  We thank Lars
Hernquist and Christine Jones for helpful discussions.  KR, AM, and MG
are supported in part by the Smithsonian Institution. JJM is supported
by Chandra Fellowship grant PF8-1003, awarded through the Chandra
Science Center.  The Chandra Science Center is operated by the
Smithsonian Astrophysical Observatory for NASA under contract
NAS8-39073.  We thank the Max-Planck Institut f\"ur Astrophysik where
some of the computing for this work was done.  This research has made
use of the NASA/IPAC 
Extragalactic Database (NED) which is operated by the Jet Propulsion
Laboratory, California Institute of Technology, under contract with
the National Aeronautics and Space Administration.

\begin{table*}[th] \footnotesize
\begin{center}
\caption{\label{properties} \sc Properties of Systems Associated with A2199}
\begin{tabular}{lcccccc}
\tableline
\tableline
\tablewidth{0pt}
System &\multicolumn{2}{c}{Coordinates} & $cz$ & $\sigma _p$ & $R_p$ & log$L_X$ \nl 
 & RA (J2000) & DEC (J2000) & $\kms$ & $\kms$ & $\Mpc$ & $h^{-2}$~ergs~s$^{-1}$ \nl 
\tableline
A2199 & 16 28 38 & 39 33 05 & 8963 &801$\pm$92$^{\tablenotemark{a}}$ & -- & 43.9 \nl
A2197W & 16 27 41 & 40 55 40 & 9300 &612$\pm$56$^{\tablenotemark{a}}$ & 2.2 & 42.9 \nl
A2197E & 16 29 43 & 40 49 12 &  --    &   --         & 2.0 & 42.7 \nl
NRGs385 & 16 17 15 & 34 55 00 & 9478 & 525$\pm$216 & 8.0 & 42.9 \nl
NRGs388 & 16 23 01 & 37 55 21 & 9788 & 468$\pm$94 & 3.1 & 42.6 \nl
\tableline
NRGs389 & 16 21 57 &  36 02 22  &   10096   & 132$\pm$38 & 5.8 & $<41.8$   \nl
NRGs396$^{\tablenotemark{b}}$ & 16 36 50 & 44 13 00 & 9540  & 480$\pm$105 & 7.6 & 42.0$^{\tablenotemark{c}}$ \nl
NRGs395 & 16 37 05   &   36 09 01 &    9957   &331$\pm$67 & 5.8  &   $<41.9$   \nl
NRGs399 & 16 41 42 &     39 47 07 &    9677   &407$\pm$94 & 3.8   &   $<41.9$   \nl
NRGs400 & 16 48 13 &     35 59 23 &    10085  &724$\pm$424 & 8.1  &   $<41.9$   \nl
\tableline
\tablenotetext{a}{Girardi et al.~1998} \nl
\tablenotetext{b}{Identified in Ramella et al.~1997; too poor to be included in
MBGR, but named using their convention}
\tablenotetext{c}{Detection significance = $2.7 \sigma$}
\end{tabular}
\end{center}
\end{table*}

\clearpage

\end{document}